\documentclass[12pt]{amsart}

\usepackage{amsmath,amssymb}

\makeatletter
\@addtoreset{equation}{section}
\makeatother

\usepackage{enumerate}

\def\res{\mathrm {res}}
\def\sn{\mathrm {sn}}
\def\cn{\mathrm {cn}}
\def\dn{\mathrm {dn}}
\def\al{\mathrm {al}}
\def\fal{\mathfrak {al}}

\def\Sym{\mathrm {Sym}}
\def\dq{ {\dot{q}}}

\def\CC{{\mathbb C}}
\def\RR{{\mathbb R}}

\newtheorem{definition}{Definition}[section]

\newtheorem{theorem}{Theorem}[section]
\newtheorem{proposition}{Proposition}[section]

\newtheorem{remark}{Remark}[section]
\newtheorem{lemma}{Lemma}[section]

\def\dfrac#1#2{{\displaystyle\frac{#1}{#2}}}

\def\book#1{\rm{#1}, }
\def\paper#1{\textit{#1}, }
\def\jour#1{\rm{#1}, }
\def\yr#1{({\rm{#1}) }}
\def\vol#1{\textbf{#1}}
\def\pages#1{\rm{#1}}

\def\publaddr#1{\rm{#1}, }
\def\publ#1{\rm{#1}, }
\def\by#1{{\rm{#1}, }}

\begin{document}

\title{Neumann system and hyperelliptic al functions}

\author{Shigeki MATSUTANI}\thanks{
\noindent 8-21-1 Higashi-Linkan Sagamihara 228-0811 JAPAN\\
\hspace{0.5 cm} e-mail: rxb01142@nifty.com}

%\date{2005, June 25}

\maketitle

\begin{abstract}
This article shows that
the Neumann dynamical system is described well
in terms of the Weierestrass hyperelliptic al functions.
\end{abstract}

%\bigskip

%\subheading{PACS numbers}:
%{\centerline{\textbf{2000 MSC: 37K20, 35Q53, 14H45, 14H70 }}}

\section{Introduction}

The Neumann dynamical system is a well-known integrable nonlinear
dynamical system, whose Lagrangian for $(q, \dq) \in \RR^{2g+2}$
is given by,
\begin{gather}
  L = \frac{1}{2}\sum_{i=1}^{g+1} \dq_i^2 -
       \frac{1}{2}\sum_{i=1}^{g+1} a_i q_i^2,
       \label{eq:lag}
\end{gather}
with a holonomic constraint,
\begin{gather}
	 \Phi(q) = 0, \quad\Phi(q):=\sum_{i=1}^{g+1} q_i^2 - 1.
	 \label{eq:Phi}
\end{gather}
This is studied well in
 frameworks of the dynamical system \cite{Mo},
of the symplectic geometry \cite{GS},
of the algebraic geometry \cite{Mu},
of the representation of the infinite Lie algebra \cite{AHP, S}.

Mumford gives explicit expressions of the Neumann system
in terms of hyperelliptic functions based upon classical and
modern hyperelliptic function theories.
This article gives more explicit expressions of the Neumann system
using  Weierstrass hyperelliptic al functions.

In the case of elliptic functions theory,
Weierstrass $\wp$ functions and Jacobi sn, cn, dn functions
play important roles in the theory
even though they are expressed by the $\theta$ functions
and all relations among them are rewritten by the $\theta$ functions.
The expressions of
Weierstrass $\wp$ functions and Jacobi sn, cn, dn functions
make the theory of elliptic functions fruitful
and reveal the essentials of elliptic functions.

Unfortunately in the case of higher genus case,
such studies are not enough though
Klein and Weierstrass defined hyperelliptic versions of
these $\wp$ functions \cite{Kl} and sn, cn, dn functions
\cite{W}. Thus several authors devote themselves
to reinterpretations of the modern theory of hyperelliptic functions
in terms of these functions and developing studies of
these functions as special functions
[\cite{BEL, Ma} and their references].
In this article, we also proceed with such a project.
We will show that the
Weierstrass al functions give natural descriptions of
the Neumann dynamical system: As in theorem 4.1,
the configuration $q^i$ of $i$-th particle (or
coordinate) is directly given by the al function,
$$
q^i(t) = \fal_i(t),
$$
Here $\fal_i(t)$ are defined in Definition 3.1,
which was originally defined by Weierstrass as
a generalization of  Jacobi
sn, cn, and dn functions over a elliptic curve
to that over a hyperelliptic curve.
 As  Jacobi sn, cn, dn
functions are associated with several nonlinear
phenomena and these relations enable us to
recognize the essentials of the phenomena \cite{T},
we expect that this expression also plays a role
in hyperelliptic function case.
In fact the description in terms of
the al functions makes several properties
of the Neumann system rather simple.
For examples, an essential property of the
Neumann system $\sum_i^g (q^i(t))^2 = 1$
is interprited as a hyperelliptic version of
$\sn^2(u) + \cn^2(u)=1$. Its hamiltonian
is given as a manifestly constant quantity
in Theorem 4.1 (3).
Due to the description, proofs in this article
basically need only primitive residual computations.

We will give our plan of this article.
\S 2 gives a short review of the Neumann system.
In \S 3, we introduce the hyperelliptic al functions
and hyperelliptic $\wp$ functions. There we
also give a short review of their basic properties
following \cite{Ba, BEL, W}.
\S 4 is our main section, where we give our
main theorem. There $\fal$ function naturally
describes the Neumann system.

\bigskip
We thank Professor Emma Previato for bringing my
attentions up on the Neumann system and Professor
Yoshihiro \^Onish for his continual supports on
the studies.

\bigskip
\section{Neumann System}

We will shortly review the Neumann system
$(q, \dq) \in \RR^{2g+2}$ whose
Lagrangian and constraint condition are
given (\ref{eq:lag}) and (\ref{eq:Phi})
in Introduction.
The constraint (\ref{eq:Phi}) means $\dot \Phi(q)=0$,
\begin{equation}
	\sum_{i=1}^{g+1} \dq_i q_i = 0. \label{eq:dPhi}
\end{equation}
The canonical momentum $p_i$ to $q_i$ is given as
$$
	p_i =\frac{\partial L}{\partial \dq_i} =  \dq.
$$

Purely kinematic investigations lead the following
proposition \cite{Mu}.

\begin{proposition}
The hamiltonian of this system is given by
\begin{gather}
           H := \frac{1}{2}\sum_{i=1}^{g+1} \dq_i^2 +
       \frac{1}{2}\sum_{i=1}^{g+1} a_i q_i^2,\label{eq:H}
\end{gather}
and the hamiltonian vector field is given by
\begin{gather}
	D_H = \sum \dq_i \frac{\partial}{\partial q_i}
	- \sum a_i q_i \frac{\partial}{\partial \dq_i}
		+ \left(\sum [a_i q_i^2-\dq_i^2]\right)
	 \sum  q_i \frac{\partial}{\partial \dq_i}.\label{eq:D}
\end{gather}
The equation of motion is given by
\begin{gather}
	\dot q_i = \dot q_i, \quad
	\ddot q_i = -(2L + a_i) q_i.  \label{eq:eq.motion}
\end{gather}
\end{proposition}

\bigskip
\section{Hyperelliptic Functions}

In this article, we will consider a hyperelliptic curve $C_g$
given by an affine equation
\cite{Mu, DRVW},
\begin{gather*}
	y^2 = f(x), \quad f(x) = A(x) Q(x),
\end{gather*}
\begin{gather*}
\begin{split}
	 A(x) &:= (x-a_1) (x-a_2) \cdots (x-a_{g+1}),\\
	 Q(x) &:= (x-c_1) (x-c_2) \cdots (x-c_{g}),
\end{split}
\end{gather*}
where $a_i$'s and $c_i$'s are complex numbers.
Let $b_i := a_i$ ($i=1, \cdots, g+1$) and $b_{g+i+1}:=c_i$
$(i=1, \cdots, g)$.

In this article, we deal with $(x_1, x_2, \cdots, x_g)$
belonging to $g$ symmetric product $\Sym^g(C_g)$
of $C_g$.

Let us introduce the canonical coordinate
$u:=(u_1, \cdots, u_g)$
in the Jacobian $\mathcal J_g$ related to $C_g$ \cite{BEL},
$$
	u_i := \sum_{a=1}^g \int^{(x_a,y_a)}_\infty \frac{x^{i-1} dx}{2y}.
$$
Here $u_-:=(u_1, \cdots, u_{g-1})$, $u=(u_-, u_g)$.

Due to Abel theorem \cite{H}, the following proposition holds.
\begin{proposition} \label{prop:u}
$(u_1, u_2, \cdots, u_g)$ are linearly independent in
$\CC^g$. In other words, there are  paths in
$\Sym^g(C_g)$ so that $\{u_g\}$ is equal to $\CC$
with fixing $u_-$.
\end{proposition}

As Mumford studied the Neumann system using
$UVW$-expression
of the hyperelliptic functions
\cite{Mu}, we will give
 $U$, $V$ and $W$ functions \cite{Mu},
\begin{gather}
U(x) =: (x-x_1)\cdots (x-x_g),
\end{gather}
\begin{gather}
	V(x) := \sum_{a=1}^g\frac{y_a U(x)}{U'(x_a) (x-x_a)}.
\end{gather}
\begin{gather}
	W(x) := \frac{f(x)+V(x)^2}{U(x)}.
\end{gather}

In this article, we will express the system in terms of
the hyperelliptic $\wp$ functions and al functions.
Let us introduce these functions,

\begin{definition}
The hyperelliptic $\wp_{g i}$ ($i = 1, 2, \cdots, g)$
functions of $u$'s are defined by
\begin{gather}
         U(x) = x^{g} + \sum_{i=1}^g (-1)^{i} \wp_{g i} x^{g-i},
\end{gather}
e.g., $\wp_{gg}:=x_1+\cdots+x_g$.

The Weierstrass $\al_i$ and $\fal_i$
($i = 1, 2, \cdots, g$) functions are
defined by \cite{Ba, W},
\begin{gather}
         \fal_r(u) := \gamma_r \al_r(u), \quad
         \al_r(u) := \sqrt{U(a_r)}(u),
\end{gather}
where we set $\gamma_r=1/\sqrt{A'(a_r)}$ in this article.
We write
$$
\al_r^{[i]}(u):=\frac{\partial}{\partial u_i} \al_r(u),
\quad
\fal_r^{[i]}(u):=\frac{\partial}{\partial u_i} \fal_r(u).
$$
\end{definition}

The sn function is defined by
$1/sn(u):= \sqrt{x - a_3}/\sqrt{A'(a_3)}$
and $\sn(u) =1/\sn(u +\Omega)$,
$\al$ functions should be recognized as
an extension of $\sn$ function.
As $\sn$ function has the relations
$$
	k^2 \sn^2(u) + \dn^2(u) =1, \quad
	\sn^2(u) + \cn^2(u) = 1.
$$
the $\fal$ functions are also has similar relations,
which were studied in \cite{Mu} as
a generalization of Frobenius identity.

\begin{proposition}\label{lemma:id}
$$
	\sum_{i=1}^{g+1}\fal_i^2(u) =1, \quad
	\sum_{i=1}^{g+1} \frac{1}{a_i}[\fal_i^{[g]}]^2(u) =1,
$$
\end{proposition}

\begin{proof}
The left hand side is given by
$$
\sum_{i=1}^{g+1}\frac{U(a_i)}{A'(a_i)} =
\frac{1}{2}\sum_{i=1}^{g+1}\res_{(a_i,0)}\frac{U(x)}{A(x)},
$$
since
around the finite ramified point $(a_i,0)$ of the curve $C_g$,
 we have
a local parameter $t^2 = (x-a_i)$ and
\begin{gather*}
\res_{(a_i,0)}\frac{U(x)}{A(x)}dx
= \res_{(a_i,0)}\frac{2 U(t^2 + a_i) t dt}
  {(t^2 + a_i - a_1) \cdots t^2\cdots (t^2 + a_i - a_{g+1}) }.
\end{gather*}
Let us consider an integral over a boundary
of polygon expression $C_0$ of $C_g$,
$$
\oint_{\partial C_0} \frac{U(x)}{A(x)} dx =0,
$$
which gives the relation,
$$
	\sum_{i=1}^{g+1}\res_{(a_i,0)}\frac{U(x)}{A(x)}dx
     =-\res_{\infty}\frac{U(x)}{A(x)}dx.
$$
At $\infty$, a local parameter $t$ of $C_g$ is given by
$x=1/t^2$:
$$
\res_{\infty}\frac{U(x)}{A(x)}dx
=\res_{\infty}
\frac{\frac{1}{t^{2g}}(1-x_1 t^2)\cdots(1-x_g t^2)}
     {\frac{1}{t^{2g+2}}(1-a_1 t^2)\cdots(1-a_{g+1} t^2)}
     \frac{-2}{t^3}dt=-2.
$$
Hence it is proved.
Similarly we obtain the relations for $\fal_r^{[g]}$
though we should evaluate $W(x)/x A(x)$.
\end{proof}

There is a natural relation between $\al$ function
and $\wp_{gg}$ function
\begin{proposition}\label{lemma:eq.motion}
$$
\frac{\partial^2}{\partial u_g^2}
\fal_i(u)=
\frac{\partial}{\partial u_g}
\fal_i^{[g]}(u)
=\left( \sum_{j=1, b_i\neq a_i}^{2g+1} b_j  -2 \wp_{gg}(u) \right)
\fal_i.
$$
\end{proposition}

\begin{proof}
This is directly obtained if we assume the following
Lemma \ref{lemma:1.2} and \ref{lemma:Miura}.
\end{proof}

We have primitive relations between
 differentials of al functions and $UVW$ expressions:
\begin{lemma}\label{lemma:1.2}
$$
	\al_i^{[g]}(u)
	= \frac{V(a_i)(u)}{ \al_i(u)}, \quad
	\fal_i^{[g]}(u)
	=\frac{V(a_i)(u)}{\fal_i(u) A'(a_i)}.
$$
\end{lemma}

\begin{proof}
Noting
$\displaystyle{\frac{\partial}{\partial u_g}=
      \sum_{a=1}^g \frac{ 2 y_a}{U'(x_a)}\frac{\partial}{\partial x_a}}
$ \cite{Ma}
and
$\displaystyle{
\frac{\partial}{\partial x_a}U(x) }$
$\displaystyle{= -\frac{U(x)}{(x-x_a)}}$,
we find that
$\displaystyle{
      \frac{1}{2}\frac{\partial}{\partial u_g}U(x)=V(x)}$,
which directly gives the relations.
\end{proof}

\begin{lemma} \label{lemma:Miura}
$$
\frac{1}{2}\frac{\partial}{\partial u_g} V(a_i)=U(a_i)
\left( \sum_{i=1}^{2g+1} b_i - a_i -2 \sum_{a=1}^{g} x_a\right)
-\frac{1}{U(a_i)}V(a_i)^2.
$$
\end{lemma}

\begin{proof}
Here we will check the left hand side.
\begin{gather*}
\begin{split}
\frac{\partial}{\partial u_g} V(a_i)
&=\sum_{a, b=1}^g\frac{y_a}{U'(x_a)}
\frac{\partial}{\partial x_a}
            \frac{y_b U(a_i)}{U'(x_b) (a_i-x_b)}\\
            &=\sum_{a=1}^g\frac{2y_a}{U'(x_a)}
\frac{\partial}{\partial x_a}
            \frac{2y_a U(a_i)}{U'(x_b) (a_i-x_b)}
+\sum_{a\neq b}^g\frac{y_a}{U'(x_a)}
\frac{\partial}{\partial x_a}
            \frac{2y_b U(a_i)}{U'(x_b) (a_i-x_b)}\\
\end{split}
\end{gather*}
\begin{gather*}
\begin{split}
&=U(a_i)\sum_{a=1}^g
\frac{1}{2}\left[\frac{1}{U'(x)}
\frac{\partial}{\partial x}\left(
            \frac{f(x) U(a_i)}{U'(x) (a_i-x_b)}\right)\right]_{x=x_a}\\
&\quad +U(a_i)\sum_{a\neq b}^g\frac{f(x_a)}{U'(x_a)^2(a_i-x_a)^2}\\
&\quad+U(a_i)\sum_{a\neq b}^g\frac{2y_a}{U'(x_a)}
            \frac{y_b}{U'(x_b)(a_i-x_a) (a_i-x_b)}\\
&=\sum_{i=1}^{2g+1}b_i - a_i - 2 \sum_{a=1}^g x_a
+U(a_i)\left(\sum_{a\neq b}^g\frac{2y_a}{U'(x_a)(a_i-x_a)}\right)^2.\\
\end{split}
\end{gather*}
Here we used the following relations.
$$
	\frac{\partial}{\partial x_a} U'(x_a) =
	\frac{1}{2}\frac{\partial}{\partial x} U(x)|_{x=x_a},
$$
$$
\left[\frac{1}{U'(x)}
\frac{\partial}{\partial x}\left(
            \frac{f(x) U(a_i)}{U'(x) (a_i-x_b)}\right)\right]_{x=x_a}
= \res_{(x_a,y_a)} \frac{f(x)}{U(x)^2 (a_i-x)}dx,
$$
and
$$
\sum_{i=1}^{2g+1}b_i - a_i - 2 \sum_{a=1}^g x_a
= \sum_{a=1}^{g}\res_{(x_a,y_a)} \frac{f(x)}{U(x)^2 (a_i-x)}dx.
$$
The third relation is obtained by an evaluation of the integral
\break
$\displaystyle{
       \oint_{\partial C_0}\frac{f(x)}{U(x)^2 (a_i-x)}dx}$.
\end{proof}

\begin{remark}{\rm
 The Klein hyperelliptic
$\wp$ function obeys the KdV equations
\cite{BEL, Ma}. On the other hand,
$\dfrac{\partial}{\partial u_g} \log \al_r$
is a solution of the MKdV
equation \cite{Ma}. The relation in
Proposition \ref{lemma:eq.motion}
means so-called {\it Miura transformation},
$$
	\left(\frac{\partial}{\partial u_g}\log \al_i\right)^2
	+\frac{\partial^2}{\partial u_g^2}\log al_i=(\mathcal L-a_i),
$$
where $\mathcal L := \dfrac{1}{2}\left( 2 \wp_{gg}
    -\sum_{i=1}^{2g+1} b_i \right)$.
}
\end{remark}

\bigskip

\section{Neumann system and hyperelliptic al functions}

This section gives our main theorem as follows.

\begin{theorem} \label{lemma:q^2}
Suppose that configurations of $(x_1, \cdots, x_g) \in \Sym^g(C_g)$
are given so that $(\al_i)$ belongs to $\RR^{g+1}$, $u_g \in \RR$ fixing
$u_- \in \RR^{g-1}$.
\begin{enumerate}

\item $\fal_i$ obey the Neumann system, i.e.,
\begin{gather}
q_i(t) = \fal_i(u_-, t), \quad\dq_i = \fal_i^{[g]}(u_-, t),
\end{gather}
 where
the time $t$ of the system is identified with $u_g$
and thus the Hamiltonian vector field is given by
$$
	D_H:=\frac{d}{dt} \equiv \frac{\partial}{\partial u_g}.
$$

\item
The hamiltonian (\ref{eq:H}) and the lagrangian
(\ref{eq:lag})  are given by
$$      H =
    \frac{1}{2}\left( \sum_{i=1}^{g+1} a_i - \sum_{a=1}^{g} c_a\right),
\quad
           L =
    \frac{1}{2}\left( 2 \wp_{gg}
    -\sum_{i=1}^{2g+1} b_i \right).
$$

\item The conserved quantities are $c_i$ ($i=1, \cdots, g$) and
$$
	m_i:= q_i^2 + \sum_{i=1}^{g+1}\sum_{j=1, \neq i}^{g+1}
             \frac{(q_i \dq_j - q_j \dq_i)^2}{a_i - a_j},
             \quad (i=1, \cdots, g+1),
$$
which obey relations,
$$
	m_i =\frac{Q(a_i)}{A'(a_i)}, \quad
	\sum_{i=1}^{g+1} m_i = 1,\quad
	\sum_{i=1}^{g+1} a_i m_i = H.
$$

\end{enumerate}
\end{theorem}

These relations were essentially proved in \cite{Mu} using
$UVW$ expression without al functions. However from the viewpoint
of studies of special functions, we will show them directly
using nature of al functions.

\begin{proof}
Assumptions are asserted by  Proposition \ref{prop:u}.
(1):
Due to Proposition \ref{lemma:id}, $\fal_r$'s
obviously obeys the constraint condition $\Phi(\fal)=0$
(\ref{eq:Phi}) and $\dot \Phi(\fal)=0$
(\ref{eq:dPhi}) by differentiating the both sides
of the identity in $u_g$. We should check whether
they obey the equation of motion (\ref{eq:eq.motion}),
which are proved in Proposition \ref{lemma:eq.motion}
if we assume the form of the Lagrangian $L$ in (2).
(2) is directly obtained by using the relations
in Lemma \ref{lemma:LH}. Finally
(3) is proved in  Remark \ref{remark:m}.
\end{proof}

\begin{remark}
{\rm
\begin{enumerate}

\item The equation of motion (\ref{eq:eq.motion})
is directly related to Proposition \ref{lemma:eq.motion},
which is connected with the Miura transformation.
Further the constraint (\ref{eq:Phi}) satisfies
due to the identity of $\fal$ function as mentioned
in Proposition \ref{lemma:id}. These exhibits essentials
of $\al$ functions. Hence the Neumann system should
be expressed by the al function as some dynamical
systems are expressed by Jacobi sn, cn, dn functions
\cite{T}.

\item
We remark that the hamiltonian depends only upon
$a_i$'s and $c_i$'s which determines the hyperelliptic
curve $C_g$. Thus it is manifest that it is invariant
for the time $u_g$ development of the system.

\item
There are $2g$ degrees of freedom as a kinematic system
because the constraints $\Phi$ and $\dot \Phi$
reduce $(2g+2)$ ones to $2g$ ones. The independent
conserved quantities $m_i$ are $g=g+1-1$;
$"-1"$ comes from $\sum m_i=1$.
Since the sum of $m_i$ gives hamiltonian $H$, $H$
is not linearly independent conserved quantities.
Since there are other $g$ conserved quantities $c_i$ but their
sum gives the hamiltonian $\sum m_i$, the
dimensional of independent $c_i$ is $g-1$.
However  $\sum_{i=1}^{g+1}\dq_i^2/a_i =1$
compensates the lacking one.
Hence the degrees of freedom of this system is
equal to number of the conserved quantities.

\item
By the definition of $c_i$'s, $c_i$ depends
upon the initial condition of the Neumann system
whereas $a_i$ is fixed as coupling constants of
the Neumann system.
Thus $\mathcal S_g:=
\{C_g\ :\ y^2 =A(x) Q(x)\ |$ $ c_1, c_2, \cdots
c_g \in \CC\}$ corresponds to the solution space $\mathcal N_g$
of the Neumann system if $u_g \in \RR$
and $(\fal, \fal^{[g]}) \in \RR^{2g+2}$.
The  $\mathcal S_g$ is
a subspace of
the moduli $\mathcal M_g$ of hyperelliptic curves of genus
$g$.

\end{enumerate}
}
\end{remark}

Let us give a lemma and remarks as follows,
which are parts of the proofs of the theorem.

\bigskip
\begin{lemma}\label{lemma:LH}
\begin{enumerate}
\item
$\displaystyle{
    \sum_{i=1}^{g+1} [\fal^{[g]}_i(u)]^2
       =\wp_{gg}(u)- \sum_{a=1}^{g} c_a.
}$

\item
$\displaystyle{
\sum a_i \fal_i(u)^2 =\sum_{i=1}^{g+1} a_i - \wp_{gg}(u).
}$

\end{enumerate}
\end{lemma}

\begin{proof} 1)
Due to Lemma \ref{lemma:1.2},
we deal with
$
\displaystyle{ \oint_{\partial C_0}\frac{V(x)^2}{U(x)A(x)} dx =0}
$
giving
$$
	2\sum_{i=1}^{g+1}  \frac{V(a_i)^2}{U(a_i)A'(a_i)}
+\sum_{a=1, \epsilon =\pm}^{g}
 \res_{(x_a, \epsilon y_a)}\frac{V(x)^2}{U(x)A(x)} dx
        +\res_{\infty}\frac{V(x)^2}{U(x)A(x)} dx =0.
$$
Whereas the third term vanishes,
each element in the second term is given by
$$
\res_{(x_a, \pm y_a)}\frac{V^2(x)}{U(x)A(x)}dx = \frac{Q(x_a)}{U'(x_a)}.
$$
Further we also evaluate an integral,
$
\displaystyle{ \oint_{\partial C_0}\frac{Q(x)}{U(x)} dx =0}.
$
The integrand has singularities at $(x_a, \pm y_a)$ and the infinity.
Similar consideration leads us to the identities
$$
\sum_{a=1, \epsilon =\pm}^{g}
  \frac{Q(x_a)}{U'(x_a)}=2(c_1 +\cdots +c_g)-2(x_1 + \cdots + x_g).
$$
Due to these relations, we have the finial equal.

2) Next we will consider an integral,
$\displaystyle{
 \oint_{\partial C_0}x\frac{U(x)}{A(x)} dx =0}$.
A residual computation gives
$\displaystyle{
	\sum_{i=1}^{g+1} a_i \frac{U(a_i)}{A'(a_i)}
        =-\res_{\infty} x\frac{U(x)}{A(x)} dx}$.
The infinity term gives
$2((x_1 + \cdots x_g) - (a_1 +\cdots a_{g+1}))$.
Hence we also have the relation in (2).
\end{proof}

\begin{remark}
Using the fact
$\displaystyle{
\frac{\partial x_a}{\partial u_g} = \frac{ 2 y_a}{U'(x_a)}}$,
we obtain anther form of Lemma \ref{lemma:LH}  \cite{DRVW},
$\displaystyle{
    \sum_{i=1}^{g+1} [\fal^{[g]}_i]^2
    = \sum_{a, b=1}^{g} g(x)_{a,b} \frac{\partial x_a}{\partial u_g}
       \frac{\partial x_b}{\partial u_g}
}$,
where
$
g(x)_{a,b}$ $:=-\displaystyle{\sum_{i=1}^{g+1}
    \frac{U(a_i)}{ (a_i - x_a)(a_i-x_b) A'(a_i)}}
$
whose off-diagonal part does not vanish
 for the case genus $g>2$ in general.
\end{remark}

\begin{remark}\label{remark:m} (Proof of Theorem 4.1 3)
Here we will give the conserved quantities of the
Neumann system.
Let us consider,
$$
	m_i(x) = q_i^2 + \sum_{i=1}^{g+1}\sum_{j=1, \neq i}^{g+1}
             \frac{(q_i \dq_j - q_j \dq_i)^2}{x - a_j}.
$$
Then we have identities
$$
	\frac{f(x)}{A(x)^2}\equiv \frac{U(x)W(x)-V(x)^2}{A(x)^2}
	=\sum_{i=1}^{g+1}\frac{m_i(x)}{x-a_i},
$$
\begin{gather*}
    m_i = \res_{a_i}\frac{m_i(x)}{x-a_i}
    =q_i^2 + \sum_{i=1}^{g+1}\sum_{j=1, \neq i}^{g+1}
            \frac{(q_i \dq_j - q_j \dq_i)^2}{a_i - a_j}.
\end{gather*}
The direct computation gives the  relations in Theorem
when we deal with the integrals of differentials
$\displaystyle{
	\frac{Q(x)}{A(x)}dx}$,
$\displaystyle{ \frac{xQ(x)}{A(x)}dx}$.
\end{remark}

\end{document}